\documentclass[11pt,twoside]{article}
\usepackage{amsfonts}
\usepackage{amsbsy}
\usepackage{amssymb}
\usepackage{amsmath}
\usepackage{amsthm}
\usepackage{mathrsfs}
\usepackage{indentfirst}
\usepackage[pdftex]{graphicx}
\usepackage{subfigure}
\usepackage{supertabular}
\usepackage{url}
\usepackage{cite}
\usepackage{hyperref}
\usepackage{color}
\definecolor{refcolor}{RGB}{0,0,190}
\hypersetup{
    colorlinks,
    citecolor=refcolor,
    filecolor=refcolor,
    linkcolor=refcolor,
    urlcolor=refcolor
}

\newtheorem{definition}{Definition}

\newtheorem{remark}{Remark}

\theoremstyle{definition}

\def\({\left(}
\def\){\right)}

\newcommand{\tn}{\textnormal}
\newcommand{\op}[1]{\tn{#1}}

\newcommand{\R}{\mathbb{R}}
\newcommand{\N}{\mathbb{N}}

\newcommand{\de}{\tn{d}}

\newcommand{\ds}{\displaystyle}

\newcommand{\ie}{{\em i.e.} }
\newcommand{\cf}{{\em cf.} }
\newcommand{\eg}{{\em e.g.} }

\newcommand{\citep}[2]{\cite{#1}, p. #2}

\newcommand{\cfeg}[2]{(\cf \eg \citep{#1}{#2})}

\newcommand{\rank}{\textnormal{rank }}

\newcommand{\dsfrac}[2]{\ds{\frac{#1}{#2}}}

\newcommand{\Ric}{\tn{Ric}}

\newcommand{\mf}[1]{\mathfrak{#1}}
\newcommand{\mc}[1]{\mathcal{#1}}
\newcommand{\ms}[1]{\mathscr{#1}}

\newcommand{\sref}[1]{\S\ref{#1}}

\newcommand{\tensors}[3]{\mc T{}^{#1}_{#2}#3}

\newcommand{\metric}[1]{\langle#1\rangle}
\newcommand{\annihg}{\coannih{g}}

\newcommand{\idxannih}[2]{#1{}^{#2}{}}
\newcommand{\idxcoannih}[2]{#1{}_{#2}{}}

\newcommand{\radix}[1]{\idxcoannih{#1}{\circ}}
\newcommand{\annih}[1]{\idxannih{#1}{\bullet}}
\newcommand{\coannih}[1]{\idxcoannih{#1}{\bullet}}

\newcommand{\kosz}{\mc K}
\newcommand{\vectmodule}{\mf X}
\newcommand{\fivect}[1]{\vectmodule(#1)}
\newcommand{\cocontr}{{{}_\bullet}}

\newcommand{\codiff}{\delta}

\newcommand{\End}{\op{End}}

\newcommand{\annihprod}[1]{\coannih{\langle\!\langle#1\rangle\!\rangle}}

\newcommand{\fivectnull}[1]{\vectmodule_\circ(#1)}
\newcommand{\fiscal}[1]{\ms F(#1)}

\newcommand{\fiformk}[2]{\mc A^{#1}(#2)}

\newcommand{\annihformsk}[2]{\annih{\mc A}{}^{#1}(#2)}
\newcommand{\discformsk}[2]{\annih{\mc A}{}^{#1}_d(#2)}

\newcommand{\srformsk}[2]{\annih{\ms A}{}^{#1}(#2)}

\newcommand{\der}{\nabla}
\newcommand{\dera}[1]{\der_{#1}}

\newcommand{\lder}{\der^{\flat}}
\newcommand{\ldera}[1]{\lder_{#1}}
\newcommand{\lderb}[2]{\ldera{#1}{#2}}
\newcommand{\lderc}[3]{(\lderb{#1}{#2})(#3)}

\newcommand{\abs}[1]{|#1|}

\def\hyph{-\penalty0\hskip0pt\relax}
\newcommand{\rstationary}{radical{\hyph}stationary}
\newcommand{\rannih}{radical{\hyph}annihilator}
\newcommand{\semiriem}{semi{\hyph}Riemannian}

\newcommand{\semireg}{semi{\hyph}regular}

\newcommand{\nondeg}{non{\hyph}degenerate}

\newcommand{\rn}{Reissner-Nordstr\"om}
\newcommand{\kn}{Kerr-Newman}

\newcommand{\CS}{\mathbb{S}}
\newcommand{\CT}{\mathbb{T}}

\begin{document}
\thispagestyle{empty}

\centerline{\large{\bf Gauge theory at singularities}}
\medskip

\centerline{\bf Ovidiu Cristinel STOICA$^{1}$}

\begin{center}
$^{1}$Department of Theoretical Physics, National Institute of Physics and Nuclear Engineering -- \textit{Horia Hulubei}, Bucharest, Romania, e-mail: cristi.stoica@theory.nipne.ro
\end{center}

\pagestyle{myheadings}
\markboth{Ovidiu Cristinel STOICA
}{Gauge theory at singularities}

\bigskip

\begin{abstract}

Building on author's previous results in singular semi-Riemannian geometry and singular general relativity, the behavior of gauge theory at singularities is analyzed. The usual formulations of the field equations at singularities are accompanied by infinities which block the evolution equations, mainly because the metric is singular, hence the usual differential operators, constructed from the metric, blow up. However, it is possible to give otherwise equivalent formulations of the Einstein, Maxwell and Yang-Mills equations, which in addition admit solutions which can be extended beyond the singularities. The main purpose of this analysis are applications to the black hole information loss paradox. An alternative approach can be made in terms of the Kaluza-Klein theory.
\\

2000 \textit{Mathematics Subject Classification:}
53A99, 
53Z05, 
83C75, 
83C05, 
83C22, 
83C50, 
70S15. 

\textit{Key words:} Singular semi-Riemannian geometry, singularities in general relativity, gauge theory, black hole information loss paradox, Maxwell equations, Einstein-Maxwell equations, Yang-Mills equations.
\end{abstract}

\section{Introduction}

\subsection{Motivation}

As it is known from the singularity theorems of Penrose and Hawking  \cite{Pen65,Haw66i,Haw66ii,Haw67iii,Pen69,HP70}, general relativity predicts, under very general conditions, the occurrence of singularities. At singularities, some of the quantities which are used in geometry and in physics become singular.

However, recent mathematical results \cite{Sto11a,Sto11b,Sto11d} showed that {\semiriem} geometry, the geometry used in general relativity, can be extended at singularities in many important cases, so that we can give descriptions containing the same geometric information, by using only non-singular geometric objects (see section \sref{s_singular_semi_riem}). These new methods were applied to the singularities in general relativity \cite{Sto13a,Sto12fqxi,Sto12b,Sto12d}, in particular to big-bang singularities \cite{Sto11h,Sto12a,Sto12c}, and to black hole singularities \cite{Sto11e,Sto11f,Sto11g,Sto12e,Sto14a}, showing that they are not as bad as initially thought, and that they even help solving other problems accompanied by infinities \cite{Sto11f,Sto11g,Sto12d}. One important result is that we can rewrite Einstein's equation so that it can be extended smoothly at singularities \cite{Sto11a,Sto12b}.

But what happens to other fields, can they be extended at singularities in a similar way? In the case of the charged stationary black holes, the answer is positive, provided that we use the proper coordinate systems \cite{Sto11f,Sto11g}. In the mentioned cases, both the potential and the electromagnetic field are analytic even at the singularity. The reason why in the usual coordinates for the {\rn} and {\kn} black holes they appeared singular is due to the fact that, in order to move from the proper coordinates to the usual ones, we have to make a singular transformation. So, we can say that the coordinates that were used before were themselves singular, and this led to a singularity of the fields which is only apparent.

This article aims to advance our understanding of the electromagnetic and Yang-Mills fields at singularities. We look first at the electromagnetic field, since, although it is abelian, it exhibits all the features which are relevant to our discussion (section \sref{s_maxwell_eq_singularities}).

The electromagnetic field $F_{ab}$ satisfies {\em Maxwell's equations}
\begin{equation}
\label{eq_maxwell}
\bigg\{
\begin{array}{lll}
\de F &=& 0, \\
\codiff F &=& J, \\
\end{array}
\end{equation}
where $J$ is a differential $1$-form representing the electric four-current. The first equation involves only the exterior differential $\de$, so it is independent on the metric. On the other hand, to define the codifferential $\codiff$, one normally needs the Hodge $\ast$ operator, which is defined only for a non-degenerate metric. One can avoid using $\codiff$, and consider instead that the exterior differential $\de$ is applied to the two-form $\ast F$, but this doesn't solve the issue, because it relies on the undefined Hodge $\ast$ operator. However, as we shall see, we can define the codifferential operator $\codiff$ without the Hodge $\ast$ operator. This allows us to rewrite the Maxwell equations in a way that is defined at singularities too and admits smooth solutions. On the other hand, solutions that are distributions are also useful, since they represent charged currents associated to point-like particles. The analysis of Maxwell's equations extends almost straightforwardly to Yang-Mills equations (section \sref{s_yang_mills_eq_singularities}).

The singular {\semiriem} manifolds satisfying the vacuum Einstein equation are the simplest and easier to understand. When other fields are involved, we can use equations which are equivalent to Einstein's outside the singulatities, but extend smoothly at singularities too \cite{Sto11a,Sto12b}. However, in vacuum, the Einstein equation is simply the condition of Ricci flatness. Therefore, if we could describe the Maxwell and Yang-Mills fields by vacuum Einstein equations, we could obtain some insights into the behavior of these fields at singularities. Fortunately, such a description can be obtained by using the Kaluza-Klein theory, and it will be discussed in section \sref{s_kaluza_klein_singulatities}. Section \sref{s_open_questions} contains some open questions.

\section{Singular manifolds}
\label{s_singular_semi_riem}

This section recalls some basic notions and results about singular manifolds, from \cite{Sto11a}, which will be used in the remainder of the article.

A {\em singular {\semiriem} manifold} $(M,g)$ is a differentiable manifold $M$ endowed with a symmetric bilinear form $g\in \Gamma(T^*M \odot_M T^*M)$ named {\em metric}  \cite{Moi40,Str41,Str42a,Str42b,Str45,Vra42,Kup96,Bej96}. This includes {\em {\semiriem} manifolds}, having the metric {\nondeg}, and in particular {\em Riemannian manifolds}, when $g$ is positive definite.

Let $(V,g)$ be a finite dimensional vector space with an inner product $g$, possibly degenerate. The totally degenerate space $\radix{V}:=V^\perp$ is called the {\em radical} of $V$. The inner product $g$ on $V$ is {\nondeg} if and only if $\radix{V}=\{0\}$.
For a singular {\semiriem} manifold $(M,g)$ we define the {\em radical of $TM$}, by $\radix{T}M=\cup_{p\in M}\radix{(T_pM)}$. Let $\fivectnull{M}$ denote the set of vector fields on $M$ for which $W_p\in\radix{(T_pM)}$. 

We define the {\em \rannih} of $(M,g)$ as the fiber bundle (with variable fiber, if $g$ doesn't have constant signature)
\begin{equation}
	\annih{T}M=\bigcup_{p\in M}\annih{(T_pM)},
\end{equation}
where $\annih{(T_pM)} \subseteq T^*_pM$ is the space of covectors at $p$ of the form $\omega_p(X_p)=\metric{Y_p,X_p}$, for some vectors $Y_p\in T_p M$ and any $X_p\in T_p M$.

We denote the $\fiscal{M}$-module of {\em radical-annihilator $k$-forms} by 
\begin{equation}
	\annihformsk k M := \Gamma\left(\bigwedge^k \annih{T}M\right),
\end{equation}
and by $\discformsk k M$ the discontinuous $k$-forms that are from $\annihformsk k M$ on the regions of constant signature.

On $\annih{T}M$ there is a unique {\nondeg} inner product $\annihg$, defined by $\annihprod{\omega,\tau}:=\annihg(\omega,\tau):=\metric{X,Y}$, where $\annih X=\omega$, $\annih Y=\tau$, $X,Y\in\fivect M$.

A tensor $T$ of type $(r,s)$ is named {\em \rannih} in the $l$-th covariant slot if  $T\in \tensors r{l-1}{M}\otimes_M\annih{T}M\otimes_M \tensors 0{s-l}{M}$. There is a unique and canonical {\em covariant contraction} or {\em covariant trace} between covariant slots which are {\rannih}, defined by the inner product $\annihg$.
For a tensor field $T$ we denote the contraction covariant $C_{kl} T$ by
\begin{equation*}
T(\omega_1,\ldots,\omega_r,v_1,\ldots,\cocontr,\ldots,\cocontr,\ldots,v_s).
\end{equation*}

For a non-degenerate metric, the covariant derivative of a vector field $Y$ in the direction of a vector field $X$, where $X,Y\in\fivect{M}$, is well defined, by the {\em Koszul formula} (see \eg \citep{ONe83}{61}). If the metric is degenerate, the covariant derivative can't be extracted from the Koszul formula. We will use instead the right part of the Koszul formula, since it remains smooth even for degenerate metrics:
\begin{equation*}
	\kosz:\fivect M^3\to\R,
\end{equation*}
\begin{equation}
\label{eq_Koszul_form}
\begin{array}{llll}
	\kosz(X,Y,Z) &:=&\ds{\frac 1 2} \{ X \metric{Y,Z} + Y \metric{Z,X} - Z \metric{X,Y} \\
	&&\ - \metric{X,[Y,Z]} + \metric{Y, [Z,X]} + \metric{Z, [X,Y]}\}.
\end{array}
\end{equation}
We call it the {\em Koszul form}. Its properties are similar to those of the covariant derivative, and were studied in \cite{Sto11a}.

Let $X,Y\in\fivect M$. The {\em lower covariant derivative} of $Y$ in the direction of $X$ is defined as the differential $1$-form $\lderb XY \in \fiformk 1{M}$
\begin{equation}
\label{eq_l_cov_der_vect}
\lderc XYZ := \kosz(X,Y,Z),
\end{equation}
for any $Z\in\fivect{M}$.
We also define the {\em lower covariant derivative operator}
\begin{equation}
	\lder:\fivect{M} \times \fivect{M} \to \fiformk 1{M},
\end{equation}
which associates to each $X,Y\in\fivect{M}$ the differential $1$-form $\ldera XY$.

A singular manifold $(M,g)$ is {\em {\rstationary}} if it satisfies the condition 
\begin{equation}
\label{eq_radical_stationary_manifold}
		\kosz(X,Y,\_)\in\annihformsk 1 M,
\end{equation}
for any $X,Y\in\fivect{M}$ (see \cite{Kup96} Definition 3.1.3).

Let $X\in\fivect{M}$, $\omega\in\annihformsk 1 {M}$, where $(M,g)$ is {\rstationary}. The covariant derivative of $\omega$ in the direction of $X$ is defined as
\begin{equation*}
	\der:\fivect{M} \times \annihformsk 1 {M} \to \discformsk 1 M,
\end{equation*}
\begin{equation}
	\left(\der_X\omega\right)(Y) := X\left(\omega(Y)\right) - \annihprod{\lderb X Y,\omega}.
\end{equation}

If the singular manifold $(M,g)$ is {\rstationary}, we define:
\begin{equation}
	\srformsk 1 M = \{\omega\in\annihformsk 1 M|(\forall X\in\fivect M)\ \der_X\omega\in\annihformsk 1  M\}.
\end{equation}

The {\em Riemann curvature tensor} is defined as
\begin{equation*}
	R: \fivect M\times \fivect M\times \fivect M\times \fivect M \to \R,
\end{equation*}
\begin{equation}
\label{eq_riemann_curvature}
	R(X,Y,Z,T) := (\dera X {\ldera Y}Z - \dera Y {\ldera X}Z - \ldera {[X,Y]}Z)(T)
\end{equation}
for any vector fields $X,Y,Z,T\in\fivect{M}$.

A singular manifold $(M,g)$ satisfying
\begin{equation}
	\ldera X Y \in\srformsk 1 M
\end{equation}
for any vector fields $X,Y\in\fivect{M}$, is called {\em {\semireg} manifold}.
A {\rstationary} manifold $(M,g)$ is {\semireg} if and only if for any $X,Y,Z,T\in\fivect M$
\begin{equation}
	\kosz(X,Y,\cocontr)\kosz(Z,T,\cocontr) \in \fiscal M.
\end{equation}

The Riemann curvature of a {\semireg} manifold $(M,g)$ is a smooth tensor field $R\in\tensors 0 4 M$.
It satisfies 
\begin{equation}
\begin{array}{lll}
	R(X,Y,Z,T) &=& X\left(\lderc Y Z T\right) - Y\left(\lderc X Z T\right) - \lderc {[X,Y]}ZT \\
&& + \annihprod{\lderb XZ,\lderb Y T} - \annihprod{\lderb YZ,\lderb X T} \\
\end{array}
\end{equation}
for any vector fields $X,Y,Z,T\in\fivect{M}$.

In a {\semireg} four dimensional spacetime, the densitized Einstein tensor $G_{ab}\det g$ is smooth \cite{Sto11a}, so a densitized version of the Einstein equation,
\begin{equation}
\label{eq_einstein:densitized}
	G\det g + \Lambda g\det g = \kappa T\det g,
\end{equation}
is smooth even at singularities,
where $\kappa:=\dsfrac{8\pi \mc G}{c^4}$, $\mc G$ is Newton's constant, and $c$ the speed of light.
In some conditions the equation is smooth even if we replace $\det g$ with $\sqrt{\det g}$.
Also, if the Riemann tensor admits a smooth Ricci decomposition, there is an alternative version of Einstein's equation, called the {\em expanded Einstein equation}, which is smooth at singularities too \cite{Sto12b}. Both these equations are equivalent to Einstein's on the regions where the metric is \nondeg.

\section{Maxwell's equations at singularities}
\label{s_maxwell_eq_singularities}

\subsection{The exterior codifferential}
\label{s_ext_codiff}

An important differential operator present in the Maxwell and Yang-Mills equations is the codifferential. This section introduces and discusses this operator at singularities.

If the metric is {\nondeg}, the {\em Hodge dual} of a differential form is defined on the space of differential $k$-forms, and valued in the space of $n-k$-forms,
\begin{equation}
	\ast:\fiformk k M \to \fiformk {n-k} M,
\end{equation}
by
\begin{equation}
\label{eq:hodge_ast}
	\(\ast\eta\)_{i_1\ldots i_{n-k}} = \frac 1 {k!}\eta^{j_1\ldots j_k}\sqrt{\abs{\det g}}\epsilon_{j_1\ldots j_k i_1\ldots i_{n-k}}.
\end{equation}
The symbol $\ast$ is called the {\em Hodge $\ast$ operator}.

If the metric is {\nondeg}, we can define the {\em codifferential} operator, which is the adjoint of the exterior derivative operator, or $\codiff$, with the help of Hodge's star operator $\ast$ \cfeg{Nak03}{250}:
\begin{equation}
	\codiff_k:\fiformk k M \to \fiformk {k-1} M
\end{equation}
by
\begin{equation}
	\codiff_k = (-1)^k\ast^{-1} \de_k \ast.
\end{equation}
When $k$ is understood, one can omit it and simply write $\codiff$ instead of $\codiff_k$.

If the metric is degenerate, but has constant signature, we can try to cook a similar definition
\begin{equation}
	\codiff_k:\annihformsk k M \to \discformsk {k-1} M,
\end{equation}
by using a Hodge $\ast$ operator defined this time at each point on the exterior algebra over the {\rannih} space:
\begin{equation}
	\ast:\annihformsk k M \to \discformsk {\rank g-k} M.
\end{equation}

If the signature of the metric is variable, then the $\ast$ operator is not continuous, in fact the dimension of the codomain space changes from point to point, as the signature changes. But the codomain of $\codiff$ remains $\wedge^{k-1}T^* M_p$ at any point $p$, so this should not be a problem. But it would be useful to find another definition of the adjoint exterior derivative operator $\codiff$. Since the domain and codomain of $\codiff$ do not change when the rank of $g$ changes, it may be possible that it acts continuously on some exterior differential forms.

\begin{remark}
A way to define the exterior codifferential on {\semiriem} manifolds, without using the duality given by the Hodge $\ast$ operator, is by the following formula
\begin{equation}
	(\codiff\omega)(X_1,\ldots,X_{k-1}) := - \dsfrac 1{(k-1)!}\sum_{a=1}^{n}\nabla_{E_a} \omega(E_a,X_1,\ldots,X_{k-1}),
\end{equation}
where $(E_a)_{a=1}^n$ is an orthonormal frame.

This suggests the following definition:
\end{remark}

\begin{definition}
Let $(M,g)$ be a {\rstationary} {\semiriem} manifold, and $\omega\in\annihformsk k M$. The {\em codifferential} of $\omega$ is the differential form $\codiff\omega$ defined by
\begin{equation}
	(\codiff\omega)(X_1,\ldots,X_{k-1}) := - \dsfrac 1{(k-1)!}\nabla_{\cocontr} \omega(\cocontr,X_1,\ldots,X_{k-1})
\end{equation}
\end{definition}
This formula can be shortened to
\begin{equation}
	\codiff\omega := - \dsfrac {i_\cocontr(\nabla_\cocontr\omega)}{(k-1)!}.
\end{equation}

The codifferential form is not guaranteed to be smooth for all $\omega\in\annihformsk k M$.



\subsection{Maxwell's equations at singularities}
\label{ss_maxwell_eq_singularities}

Section \sref{s_ext_codiff} introduced a way to construct it which avoids the use of the $\ast$ operator, which in the case of a differential $2$-form $F$ takes the expression
\begin{equation}
\label{eq_codiff_FX}
	(\codiff F)(X) := - \nabla_{\cocontr} F(\cocontr,X),
\end{equation}
or
\begin{equation}
\label{eq_codiff_F}
	\codiff F := - i_\cocontr(\nabla_\cocontr F),
\end{equation}
or
\begin{equation}
\label{eq_codiff_F_coords}
	\(\codiff F\)_a := - \nabla_{\cocontr} F_{\cocontr a}.
\end{equation}

It is straightforward now to generalize Maxwell's equations at singularities -- they are formally just equations \eqref{eq_maxwell}, with the specification that the codifferential operator from the second equation is defined as in equation \eqref{eq_codiff_FX}. In the case when the metric is non-degenerate, the equations coincide with the usual ones.

It is not always needed that the solution to the equation \eqref{eq_codiff_FX} is smooth: a distribution solution is good enough, since the source may be for example a point-like particle (see section \sref{s_maxwell_example_rn}).

\subsection{Example: the analytic {\rn} solution}
\label{s_maxwell_example_rn}

The {\rn} solution to the Einstein-Maxwell equations, describing a static, spherically symmetric, electrically charged, non-rota\-ting black hole is \cite{reiss16,nord18,HE95}
\begin{equation}
\label{eq_rn_metric}
\de s^2 = -\left(1-\dsfrac{2m}{r} + \dsfrac{q^2}{r^2}\right)\de t^2 + \left(1-\dsfrac{2m}{r} + \dsfrac{q^2}{r^2}\right)^{-1}\de r^2 + r^2\de\sigma^2,
\end{equation}
where $q$ and $m$ are the electric charge and the mass of the body, the units are chosen so that $c=1$ and $G=1$, and
\begin{equation}
\label{eq_sphere}
\de\sigma^2 = \de\theta^2 + \sin^2\theta \de \phi^2.
\end{equation}

We change the coordinates $r$ and $t$ in a neighborhood $r\in[0,M)$ of the singularity, where $M$ depends on whether the black hole is naked, by
\begin{equation}
\label{eq_coordinate_ext_ext}
\begin{array}{l}
\left\{
\begin{array}{ll}
t &= \tau\rho^\CT \\
r &= \rho^\CS \\
\end{array}
\right..
\\
\end{array}
\end{equation}

To make the metric analytic even at the singularity $r=0$, we choose $\CS,\CT\in\N$ given by
\begin{equation}
\label{eq_metric_smooth_cond}
\begin{array}{l}
\left\{
\begin{array}{ll}
\CS \geq 1 \\
\CT \geq \CS + 1
\end{array}
\right..
\\
\end{array}
\end{equation}

In the new coordinates, the {\rn} metric takes the form
\begin{equation}
\label{eq_rn_ext_ext}
\de s^2 = - \Delta\rho^{2\CT-2\CS-2}\left(\rho\de\tau + \CT\tau\de\rho\right)^2 + \dsfrac{\CS^2}{\Delta}\rho^{4\CS-2}\de\rho^2 + \rho^{2\CS}\de\sigma^2,
\end{equation}
where
\begin{equation}
	\Delta := r^2 - 2m r + q^2 \tn{ (hence $\Delta = \rho^{2\CS} - 2m \rho^{\CS} + q^2$)}.
\end{equation}

In the standard {\rn} solution, in coordinates $(t,r,\phi,\theta)$, the potential of the electromagnetic field is 
\begin{equation}
A = -\dsfrac q r \de t,
\end{equation}
and is singular at $r=0$.
But in the new coordinates $(\tau,\rho,\phi,\theta)$, the electromagnetic potential is
\begin{equation}
\label{eq:elmag_potential}
A = -q\rho^{\CT-\CS-1}\left(\rho\de\tau + \CT\tau\de\rho\right),
\end{equation}
and the electromagnetic field is
\begin{equation}
\label{eq:elmag_field}
F = q(2\CT-\CS)\rho^{\CT-\CS-1}\de\tau \wedge\de\rho.
\end{equation}
Both are finite, and analytic everywhere, including at the singularity $\rho=0$.

The second Maxwell equation $\codiff F = J$ gives the current density, which is a distribution vanishing outside the singularity $\rho=0$.

\section{The Yang-Mills equations at singularities}
\label{s_yang_mills_eq_singularities}

In this section, the remarks regarding Maxwell's equations at singularities are extended to the non-commutative gauge theories.

\subsection{The Yang-Mills equations}
\label{ss_yang_mills_eq}

Let $E\to M$ be a vector bundle over a manifold $M$, and $D$ a connection on $E$. Let's denote by $D_Xs$ the {\em covariant derivative} of a section $s\in\Gamma(E)$ in the direction of a vector field $X$ on $M$. If $(x_a)_{a=1}^n$ are coordinates on $U\subseteq M$, and $(e_i)_{i=1}^{\dim E}$ is a basis of local sections of $E\to U$, then the covariant derivative of a section $s$ has the components
\begin{equation}
\label{eq_gauge_cov_deriv_components}
\(D_{a}s\)^i = \partial_a s^i + A^i_{aj}s^j.
\end{equation}
The functions $A^i_{aj}$ are the components of a vector potential $A=A^i_{aj}e^j\otimes e_i\otimes \de x^a$, which is an $\End(E)$-valued $1$-form. We also denote $A_a:=A^i_{aj}e^j\otimes e_i$, and $A(X):=A_aX^a$. Since $A$ can be written as a linear combination $A=\sum_\alpha T_\alpha\otimes \omega_\alpha$, where $T_\alpha$ are sections of $\End(E)$ and $\omega_\alpha$ are $1$-forms on $M$, it follows that $A(X)=\sum_\alpha\omega_\alpha(X) T_\alpha$ is a section of $\End(E)$.

The {\em exterior covariant derivative} is defined for a section $s\in\Gamma(E)$ as
\begin{equation}
\label{eq_exterior_covariant_derivative_E}
\de_D s(X) := D_x s,
\end{equation}
and for an $E$-valued differential form 
\begin{equation}
\label{eq_exterior_covariant_derivative_E_diff}
\de_D(s\otimes \omega) := \de_D s \wedge \omega + s \otimes \de \omega,
\end{equation}
where $(s\otimes\omega)\wedge\tau := s\otimes(\omega\wedge\tau)$.

The {\em curvature} of the connection $D$ is defined as
\begin{equation}
\label{eq_gauge_curvature}
F(X,Y)s := D_X D_Y s -D_Y D_X s - D_{[X,Y]} s.
\end{equation}
Let $F_{ab}:=F(\partial_a,\partial_b)$. Then
\begin{equation}
\label{eq_gauge_curvature_components_2}
F_{abi}{}^j = \partial_a A_{bi}{}^j - \partial_b A_{ai}{}^j + A_{ak}{}^j A_{bi}{}^k - A_{bk}{}^j A_{ai}{}^k,
\end{equation}
or
\begin{equation}
\label{eq_gauge_curvature_components}
F_{ab} = \partial_a A_b - \partial_b A_a + [A_a,A_b].
\end{equation}

The curvature $F$ satisfies {\em the Bianchi identity}
\begin{equation}
\label{eq_gauge_bianchi}
\de_D F = 0,
\end{equation}
which is the first of the two Yang-Mills equations.

Let's define the Hodge $\ast$ operator on $\End(E)$-valued differential forms by
\begin{equation}
\label{eq_hodge_gauge}
\ast\(T\otimes\omega\) := T\otimes\ast\omega,
\end{equation}
and the codifferential $\codiff_D$ associated to the connection $D$ by
\begin{equation}
\label{eq_codiff_gauge}
\codiff_D := \ast \de_D \ast.
\end{equation}

Then, the {\em Yang-Mills equations} are
\begin{equation}
\label{eq_yang_mills}
\bigg\{
\begin{array}{lll}
\de_D F &=& 0, \\
\codiff_D F &=& J, \\
\end{array}
\end{equation}
where $J$ is the {\em current}.

\subsection{The Yang-Mills equations at singularities}
\label{ss_yang_mills_eq_singularities}

But what are the Yang-Mills equations at singularities?
From equations \eqref{eq_exterior_covariant_derivative_E_diff} and \eqref{eq_codiff_gauge}, it is immediate that the codifferential $\codiff_D$ associated to the connection $D$ depends only on $\de_D$ and $\de$,
\begin{equation}
\codiff_D(s\otimes \omega) = \ast\de_D\ast(s\otimes \omega) = \de_D s \wedge \ast^2\omega + s \otimes \ast\de \ast\omega = (-1)^q\de_D s \wedge \omega + s \otimes \codiff\omega
\end{equation}
where $q$ is an integer depending on the signature of the metric, the degree of the form $\omega$, and the dimension. Hence, ultimately, all that may be affected by the singularity in the second Yang-Mills equation is $\codiff F_i{}^j$, where $F_i{}^j = F_{abi}{}^j\de x^a\wedge\de x^b$. In other words, at singularities, the components of the gauge curvature behave similarly to the simpler case of Maxwell's equations.

\section{Kaluza-Klein theory with singularities}
\label{s_kaluza_klein_singulatities}

\subsection{Kaluza-Klein theory}
\label{ss_kaluza_klein}

Maxwell's theory is a gauge theory, and can be described using the fiber bundle approach. The electromagnetic potential is a connection on a principal $\op{U}(1)$-bundle $P_{\op{U}(1)}\to M$, and the electromagnetic field is the curvature of that connection. The base manifold $M$ of the principal $\op{U}(1)$-bundle is the spacetime. But why would we have two different manifolds, the five-dimensional $\op{U}(1)$-bundle manifold $P_{\op{U}(1)}$, and the four-dimensional spacetime $M$? The specification of the base manifold is redundant, since it can be obtained by factoring the manifold $P_{\op{U}(1)}$ in the fiber direction. This shows that gauge theory is, from geometric viewpoint, a Kaluza-Klein theory, with symmetries constrained by the gauge invariance.

The {\em Kaluza-Klein theory} is based on an idea of Nordstr{\"o}m \cite{nordstrom1914moglichkeit}, who proposed it in his metric theory of gravity. The general relativistic version was proposed by Kaluza \cite{kal:1921}. Because at that time the notion of a principal bundle was not yet understood, it was considered that the fifth dimension should be observable, which would contradict our experience. To hide it, Klein proposed that the fifth dimension was compact and very small \cite{kle:1926}.

In the Kaluza-Klein theory, electromagnetism is obtained as a curvature effect in the fifth dimension. More precisely, the five-dimensional metric is
\begin{equation}
	g^{(5)}_{AB} =
	\left(
\begin{array}{cc}
	g_{ab} - \xi^2 A_a A_b \phi & \xi A_a \phi \\
	\xi A_b \phi &  -\phi \\
\end{array}
\right),
\end{equation}
where $\xi^2=16\pi G$, and $\phi$ is a scalar field that, to obtain the same predictions as electromagnetism, should be constant.

By imposing the condition that $g^{(5)}_{AB}$ satisfies the vacuum Einstein equation $\Ric(g^{(5)})=0$, \ie that the five-dimenisonal manifold is Ricci flat, one obtains the {\em Einstein-Maxwell equations}, that is, the source-free Maxwell equations, and the Einstein equation for the four-dimensional metric $g_{ab}$ with the stress-energy tensor 
\begin{equation}
\label{eq_stress_energy_maxwell}
	T_{ab} = \dsfrac{1}{\mu_0}\left(F_{as}F_{b}{}^s - \dsfrac 1 4 F_{st} F^{st}g_{ab}\right)
\end{equation}
sourced by the electromagnetic field.

\subsection{Kaluza-Klein theory with singularities}
\label{ss_kaluza_klein_singularities}

We are interested now in the case when singularities are present. If the five-dimensional manifold $P_{\op{U}(1)}$ is {\semireg}, its Riemann curvature tensor $R^{(5)}_{ABCD}$ is smooth and radical-annihilator, and the Ricci tensor $R^{(5)}_{AB}=\Ric(g^{(5)})$ is well-defined at the points where the signature of the metric doesn't change. The condition that the vacuum Einstein equation is satisfied implies in particular that the Einstein tensor $G^{(5)}_{AB}=R^{(5)}_{AB}=0$ is well-defined and smooth everywhere, even at singularites.

Not the same can be said about the base manifold $M$, for which the Einstein-Maxwell equations may be singular when viewed as four-dimensional equations, although they are equivalent to smooth five-dimensional vacuum Einstein equations. In the presence of singularities, the index raising doesn't work, and we have to use the covariant contraction, the generalizations of covariant derivative and the Riemann curvature defined in section \sref{s_singular_semi_riem}. In this case, one can use equations equivalent to Einstein's, but which work at singularities too, such as the {\em densitized Einstein equation} \cite{Sto11a}, or the {\em expanded Einstein equation} \cite{Sto12a,Sto12b,Sto12d}. The Maxwell and Yang-Mills equations have to be replaced by their versions in sections \sref{s_maxwell_eq_singularities} and \sref{s_yang_mills_eq_singularities}.
The stress-energy tensor of an electromagnetic field becomes
\begin{equation}
\label{eq_stress_energy_maxwell_down}
	T_{ab} = \dsfrac{1}{\mu_0}\left(F_{a\cocontr}F_{b\cocontr} - \dsfrac 1 4 F_{\cocontr \cocontr '} F_{\cocontr \cocontr '}g_{ab}\right).
\end{equation}
It is smooth where the signature is constant, and where the signature changes, it is not smooth or even diverges. But it can be made smooth and non-singular if additional boundary conditions are imposed to $F$ where the signature changes. On the other hand, for a {\semireg} metric, $T_{ab}\det g$ is smooth even at singularities.

In conclusion, the {\semireg} spacetimes that are Ricci flat are simpler than the general ones, which need special alternatives to the Einstein equation and the other field equations. Increasing the dimension to avoid these difficulties may be fruitful, because all that remains is the vacuum Einstein equation at singularities, which is simpler.
The Kaluza-Klein theory can be extended to contain singularities in a simpler and safer way than the four-dimensional theory containing the corresponding electromagnetic fields. Moreover, the sources of the fields can be included at singularities, as we have seen in section \sref{s_maxwell_example_rn}.

\section{Some open questions}
\label{s_open_questions}

As any young fields, singular {\semiriem} geometry and singular general relativity are expected to have many yet unsolved research problems. In what concerns the issues of  gauge theory at singularities, I will mention a few open questions.
Some of them concern the relation between the gauge theory at singularities (sections \sref{s_maxwell_eq_singularities} and \sref{s_yang_mills_eq_singularities}) and the Kaluza-Klein approach (section \sref{s_kaluza_klein_singulatities}). Are the two approaches equivalent? Can we learn more about singularities in four dimensions, from singularities in Ricci flat higher dimensional spaces? What is the physical meaning of the singularities in gauge theory? Do these approaches allow matter to survive passing through the black hole's singularities during the Hawking evaporation, leading to a positive resolution of the black hole information paradox? Can these approaches reveal more about the renormalizability in quantum field theory and quantum gravity, by using dimensional reduction at singularities \cite{Sto12d}?
These are just a few open questions that, if solved, may lead to significant progress in the issues surrounding the singularities in general relativity.

\section{Conclusions}
\label{s_conclusions}

In this article, I built the case for the importance of understanding how gauge fields behave at singularities. The motivation is multiple: we need to understand their behavior at the big-bang singularity, we need to understand what happens with the fields carrying information at singularities, to see whether the information is indeed lost, we need to understand what is happening at high energy scales, and how point-like particles which can be modeled as very small black holes behave. The geometric apparatus is already in place from previous work, and it was applied to various problems involving black hole and big-bang singularities.

In this article, the methods of singular {\semiriem} geometry and singular general relativity were applied also to gauge fields on singular spacetimes. The Maxwell and Yang-Mills equations were written in a way that works at singularities too.

A particularly promising direction comes from the Kaluza-Klein theory. Since in the higher-dimensional manifolds on which Kaluza-Klein theories are defined the Einstein equation satisfies the vacuum condition, the singularities are simpler, and don't need special conditions as in the case when matter is present. This may simplify the conditions to be satisfied by the gauge fields at singularities.

These new directions are only at the beginning, since there are more implications to be explored. The aim of this article is to establish a foundation. There are some important open problems, some of them mentioned in section \sref{s_open_questions}.


\end{document}